\documentclass[preprint]{aastex}
\shortauthors{Xiang et al.}
\usepackage{multirow}
\usepackage{epsfig}
\usepackage{subfigure}
\usepackage{graphicx,times}
\input{epsf.sty}                        %for PS/EPS graphics inclusion, old

\begin{document}
%\linenumbers
\title{Study on the Temporal Evolution of the Radial Differential Rotation of Solar Corona Using Radio Emissions}
\author{N. B. Xiang\altaffilmark{1,2,3}, X. H. Zhao\altaffilmark{2}, L. H. Deng\altaffilmark{4}, F. Y. Li\altaffilmark{5,2}, Y. J. Wang\altaffilmark{4}, X. W. Tan\altaffilmark{4}}
\affil{$^{1}$Yunnan Observatories, Chinese Academy of Sciences, Kunming 650216, China; nanbin@ynao.ac.cn}
 \affil{$^{2}$ State Key Laboratory of Space Weather, National Space Science Center, Chinese Academy of Sciences, Beijing 100190, China; xhzhao@spaceweather.ac.cn}
 \affil{$^{3}$Yunnan Key Laboratory of Solar Physics and Space Science, 650216, China}
\affil{$^{4}$School of Mathematics and Computer Science, Yunnan Minzu University, Kunming 650504, China; linhua.deng@ymu.edu.cn}
 \affil{$^{5}$Institute of Optics and Electronics, Chinese Academy of Sciences, Chengdu 610209, China}
%\email{nanbin@ynao.ac.cn}

%\altaffiltext{1}{Yunnan Observatories, CAS, Kunming 650011, China}
%\altaffiltext{2}{ University of Chinese Academy of Sciences, Beijing 100049}

\begin{abstract}
The daily measurements of the disc-integrated solar radio flux, observed by the Radio Solar Telescope Network (RSTN), at 245, 410, 610, 1415, 2695, 4995, and 8800 MHz during the time interval of 1989 January 1 to 2019 December 17, are used to investigate the  temporal evolution of radial differential rotation of solar corona using the methods of Ensemble Empirical Mode Decomposition and wavelet analysis. Overall, the results reveal that over the 30-year period, the rotation rates for the observed solar radio flux within the frequency range of 245\textendash8800 MHz show  an increase with frequency. This verifies the existence of the radial differential rotation of the solar corona over long timescales of nearly 3 solar cycles. Based on the radio emission mechanism, to some extent, the results can also serve as an indicator of how the rotation of the solar upper atmosphere varies with altitude within a specific range. From the temporal variation of rotation cycle lengths of radio flux, the coronal rotation at different altitudes from the low corona to  approximately 1.3 $R_{\odot}$ exhibits complex temporal variations with the progression of the solar cycle. However, in this altitude range, over the past 30 years from 1989 to 2019, the coronal rotation consistently becomes gradually slower as the altitude increases. Finally, the EEMD method can extract rotation cycle signals from these highly randomized radio emissions, and so it can be used to investigate the rotation periods for the radio emissions at higher or lower frequencies.
\end{abstract}


\section{Introduction}
The rotation of the solar corona is a challenging but crucial study topic in solar physics. On one hand, the solar corona exhibits radial and latitudinal  differential rotation (Vats et al. 2001; Braj{\v{s}}a et al. 2002). Such coronal rotation facilitates the occurrence of magnetic reconnection in the solar corona, subsequently releasing the free energy stored in the magnetic field through shearing and twisting motions stemming from the different rotation rates of the solar surface (Giordano \& Mancuso 2008;  Zweibel et al. 2009, Li et al. 2018). This is a key to understanding the dynamics and  heating of the solar corona. On the other hand, the large-scale magnetic structures of the corona may be anchored below the photosphere, reflecting the rotation of the solar interior to some extent (Hiremath \& Hegde 2013; Mancuso et al. 2020). However, the corona, which is optically thin to extreme ultraviolet radiation, lacks virtually any long-lived solar structures (tracers) capable of displaying prominent rotating tracer features in terms of both spatial and temporal extent within it (Howard 1984; Bhatt et al.2017; Mancuso et al. 2020). Meanwhile, there is still no accurate measurement of the coronal magnetic field up to now (Bhatt et al.2017; Deng et al. 2020; Xiang et al. 2023). Thus, studying the rotation of the solar corona in a manner similar to accurately investigating the rotation of the photosphere using tracers such as sunspots is impossible. As a consequence, there were some significant differences in the results regarding the coronal rotation in previous studies (Braj{\v{s}}a et al. 2004; Giordano \& Mancuso 2008; Mancuso \& Giordano 2011; Sharma et al. 2020, 2023; Edwards et al. 2022). So far, there is still a lack of conclusive results on the coronal rotation, especially the radial differential rotation of solar corona. However, solar radio emissions are capable of revealing a wealth of information about solar coronal activities and are widely applied in various studies of the solar corona, including coronal rotation. These studies are of great significance for understanding coronal activities, the structure of the coronal atmosphere, and phenomena such as solar radio bursts (Vats et al. 2001; Singh et al. 2021; Alissandrakis et al. 2022; Li et al. 2024).

In the past three decades, different observational data were used to study the coronal rotation at various altitudes. Gawronska and Borkowski (1996) utilized daily radio flux data at 127 MHz, while Vats et al. (1998) and Chandra et al. (2010) employed daily radio flux data at 2800 MHz. Braj{\v{s}}a et al. (2004) and  Karachik et al. (2006)   analyzed coronal bright points from SOHO/EIT images. Dorotovi{\v{c}} and Rybansk{\'y}(2019) studied coronal rotation periods using bright features from SDO/AIA 21.1 nm EUV images. Additionally, Sharma et al. (2023) investigated the coronal rotation period using high-resolution images obtained from SDO/AIA at 19.3 nm. These various studies provided insights into the coronal rotation variations at different altitudes. However, the results in these studies did not address the variation of the coronal rotation period with altitude, namely, the radial differential rotation of the corona.

Early studies have indicated that radio emissions at different frequencies are generated at various altitudes in the solar corona. Based on the coronal electron density model, the original altitudes for radio emissions in the corona can be estimated, and a correspondence between radio emissions of different frequencies and altitudes can be established (Aschwanden \& Benz 1995; Vats et al. 2001). Although the original altitude for a certain radio frequency remains an approximation, it is certain that the higher the frequency of the emission, the lower the altitude of its origin tends to be. Observational data from radio emissions, therefore, offers the possibility to study the radial differential rotation of the solar corona.  Based on the analysis of the disc-integrated solar radio flux at 11 frequencies in the range of 275\textendash2800 MHz from 1997 July  to  1999 July, Vats et al. (2001) studied the variation of coronal rotation with altitude.  The authors first reported that the coronal rotation period increases with increasing frequency. The result in this study indicates that the coronal rotation accelerates as the altitude increases from the lower corona to approximately 1.2 $R_{\odot}$. Using the same radio emissions
within the same time interval in Vats et al. (2001), Bhatt et al. (2017) investigated this issue once again and obtained the opposite results: the coronal rotation period decreases with increasing frequency. The authors believed that their research results are more accurate, which is attributed to the fact that Gaussian fitting of the first secondary maximum of the autocorrelation coefficient can obtain a more accurate rotation period. Recently, Singh et al. (2021) re-investigated this issue by adding 4 additional frequencies of radio fluxes to the data used in the two mentioned above studies. They stated that the coronal rotation slows down as the altitude increases based on the data from the same time interval (1997 July  to  1999 July). However, when considering the data from 1994 to 1996, there is no systematic relationship between the variation of coronal rotation and radio emissions.

To sum up, the results on the radial differential rotation of the corona  in the previous studies are contradictory to each other, and a conclusive results is still needed regarding this issue. Additionally, the temporal variation of radial differential rotation of solar corona  as the solar cycle progressed  remains unknown. The observation of Radio Solar Telescope Network (RSTN) Noontime Flux began in 1966 May, and there have been decades of radio flux data records so far. We use these radio flux data to investigate  the temporal variation of radial differential rotation of solar corona during decades in this study, and try to obtain conclusive results regarding the variation of coronal rotation with altitude.

\section{Data and Methods}
The daily measurements of the disc-integrated solar radio flux, observed by RSTN in the frequency range of 245 MHz to 15.4 MHz, began in 1966 May. Radio flux Data spanning from 1966 May to 1987 December exclusively originate from Sagamore Hill (SGMR) in Massachusetts, and have undergone rigorous quality control measures. Since 1988, radio flux data are derived from Palehua (PALE) Hawaii, San Vito (SVTO) Italy, Learmonth (LEAR)  Australia, as well as Sagamore Hill. However, it is important to note that these data have not undergone quality control, and therefore, we only use the SGMR data in this study. For the SGMR data, not all frequencies of radio flux were measured from 1966, and there existed a large number of data gaps in the early period. In order to investigate the temporal variation of radial differential rotation of solar corona over decades, we select the daily measurements of solar radio flux at 245, 410, 610, 1415, 2695, 4995, and 8800 MHz during the time interval of 1989 January 1 to 2019 December 17. These time series of radio flux data can be downloaded from the web site https://www.ncei.noaa.gov/products/space-weather/legacy-data/solar-radio-datasets. The randomization is lowest for radio flux  near 2800 MHz, and thus the radio flux emission near this frequency is the most appropriate for investigating the coronal rotation (Vats et al. 1997; Mehta 2005; Chandra \& Vats 2011; Bhatt et al. 2018). The irregularities of the plasma in the corona and the Earth's upper atmosphere should have a strong effect on lower frequencies, and the higher frequencies are strongly influenced by the more variable activities of the inner corona (Vats et al. 1997; Bhatt et al.2017, 2018). Three typical examples of the daily radio flux at 245, 2695 and 8800 MHz  during time interval considered are shown in Figure 1. As figure 1 shows, for solar radio flux, the modulation of the solar cycle is clearly visible. However, this figure is too compressed to show the variation due to solar rotation, and this issue will be discussed in section 3.1.

\begin{figure*}[!h]
\begin{center}
\includegraphics[width=1.0\textwidth]{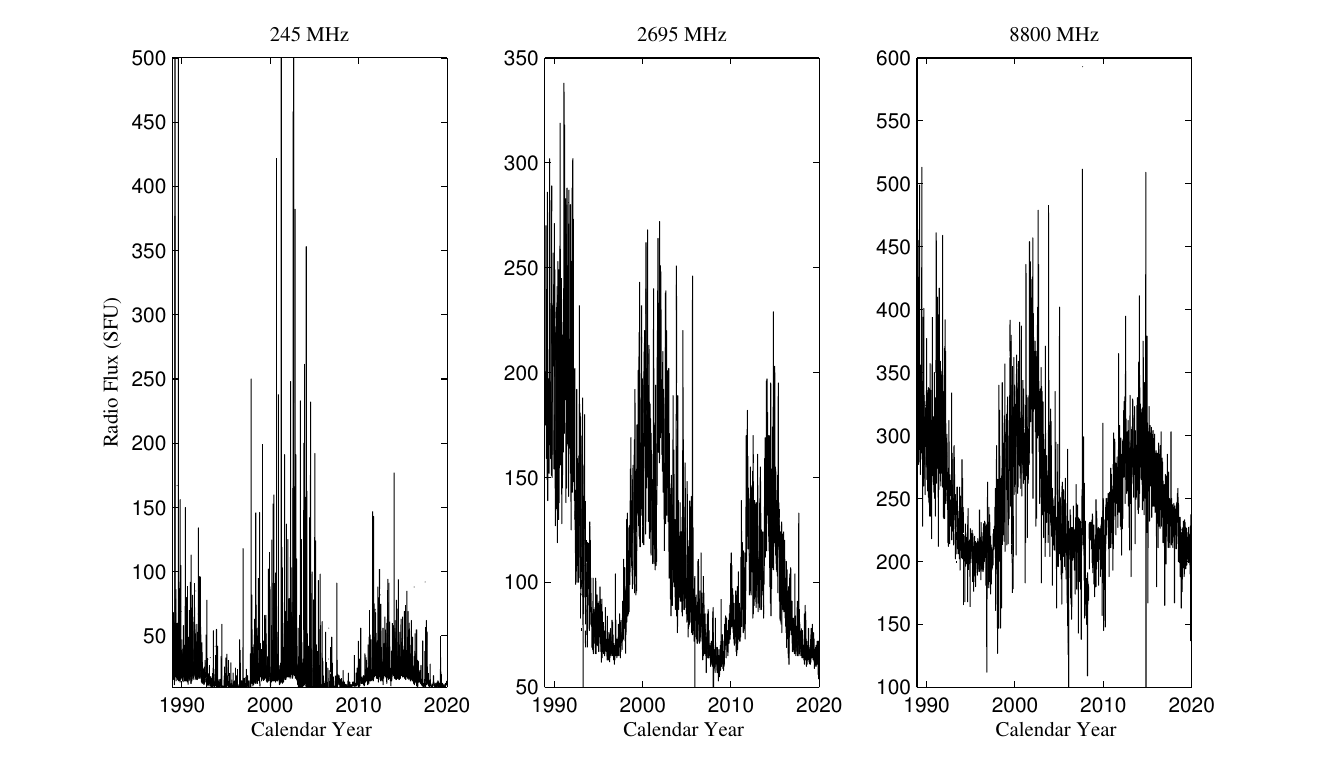}
\end{center}
\caption{Three typical examples of daily solar radio flux at 245, 2695 and 8800 MHz from 1989 January 1 to 2019 December 17 ranked from left to right, respectively.}
\end{figure*}

The Empirical Mode Decomposition (EMD) is a nonlinear time-frequency analysis algorithm, which can use Hilbert transforms to decompose complex data sets into a finite and often small number of intrinsic mode functions (IMFs).  These IMFs possess well-behaved Hilbert transforms, making them capable of accurately representing the intrinsic periodicities of the original signal (Huang et al. 1998; Barnhart \& Eichinger 2011; Xiang \& Qu 2016). However, a significant disadvantage of the EMD lies in the recurrent occurrence of mode mixing. This not only gives rise to significant aliasing in the time-frequency distribution but also obfuscates the physical meaning of individual IMF (Wu \& Huang 2009). Based on EMD, Wu and Huang (2009) proposed a new noise-assisted analysis method known as Ensemble Empirical Mode Decomposition (EEMD). The main analysis process of the EEMD method is as follows: (1)  Incorporate a set of white noise into the analyzed data; (2) Using the method of EMD, the data with white noise incorporated is decomposed, resulting in a set of IMFs. (3) Repeatedly perform steps 1 and 2, but each time, the white noise  incorporated into the analyzed data is different and uncorrelated. (4) A set of IMFs can be obtained through the preceding steps 1, 2, and 3. By averaging the corresponding IMFs within this set, the final result of the EEMD analysis is achieved. The number of times steps 1, 2, and 3 are repeated in the EEMD analysis is referred to as the number of ensemble members, i.e., ensemble size. In general, an ensemble size of a few hundred can lead to a very good result (Wu \& Huang 2009). The incorporation of white noise can establish a consistent reference framework in the time-frequency domain, enabling the compilation of signal segments with comparable scales into a single IMF (Wu \& Huang 2009; Xiang \& Qu 2016). Consequently, the EEMD solves the major drawback of the frequent appearance of mode mixing when using EMD, and a single IMF only represents an intrinsic periodicities of the original signal at a particular time scale. The EEMD method is highly suitable for extracting rotation cycle signals from the solar radio flux data.

The wavelet analysis was widely used to detect the periods of various solar activity indicators, as well as their temporal variation (Chowdhury \& Dwivedi 2011; Gurgenashvili et al. 2016, 2017; Xiang et al. 2023). For the reason, the wavelet analysis, utilizing power spectra analysis, can uncover the prominent periods at different timescales and the localized oscillatory feature  of a one-dimensional time series. This process involves decomposing the time series into a two-dimensional time-frequency space. Furthermore, the local wavelet power spectra provide detailed insights into the time-frequency components of the analyzed data, allowing for the identification of frequency components at specific points in time within the examined time series (Torrence \& Compo 1998; Grinsted et al. 2004). It is highly reasonable to choose wavelet analysis to study the temporal evolution of radial differential rotation of solar corona. In this study, we select the continuous wavelet transform (CWT), and the Morlet wave, which is used as the mother wavelet in the process of the CWT. This is because the Morlet wavelet can not only be used for feature extraction but also provides a reasonable localization in both time and frequency (Torrence \& Compo 1998; Li et al. 2009; Chowdhury \& Dwivedi 2011). For the Morlet wavelet, it is  defined as
\begin{equation}
\Psi_{0}(\eta)=\pi^{-1/4}e^{i\omega_{0}\eta}e^{-\eta^{2}/2}
\end{equation}
where $\eta$ and $\omega_{0}$ are dimensionless time and dimensionless frequency, respectively. In the continuous wavelet transform process, the oscillations within the mother wavelet are regulated by $\omega_{0}$, and thus  the choice of $\omega_{0}$ must affect the time and frequency resolution of the corresponding transform (De Moortel et al. 2004). Moreover, selecting smaller values of $\omega_{0}$ in the continuous wavelet transform leads to enhanced time resolution, whereas opting for larger values of $\omega_{0}$  facilitates achieving a superior frequency resolution, ultimately pointing towards a more precise resolution of the period (Grinsted et al. 2004; Chowdhury et al. 2010). In this study, we experimented with various values for $\omega_{0}$ and discovered that setting the dimensionless frequency $\omega_{0}=12$ presents an optimal selection, as it effectively balances time and frequency resolution capabilities. Finally, the statistical significance test of wavelet power is evaluated based on the assumption that the noise exhibits a red spectrum. To ensure the reliability of the periods measured from the wavelet analysis, the statistical significance level is  at 99.9\%.

\section{Temporal Evolution of Radial Differential Rotation of Solar Corona}
\subsection{EEMD analysis of solar radio flux data}
 The daily measurements of solar flux at 7 radio frequencies in the range of 245\textendash8800 MHz span 11,308 days, but the number of no-data days during the time interval we considered is about 433\textendash707 days, accounting for about 3.83\textendash6.25\%. In order to employ the EEMD method to decompose the daily radio flux data, and obtain the IMFs that represent the signals of the solar rotation cycle, the linear interpolation was employed to estimate the missing value in the radio flux data when there is no observed value for a certain day. The data gaps within the time interval considered are randomly distributed, and generally, the continuous gaps in these time series of solar radio flux do not exceed 6 days, significantly shorter than the timescale of a solar rotation period. Thus, the data gaps have little impact on the application of the EEMD for extracting rotation period signals from the solar radio flux data after data gap interpolation.

The EEMD is  utilized to decompose the data of 7 radio frequency fluxes. In the EEMD, an ensemble size of 500 is used, and the added white noise in each ensemble member has a standard deviation of 0.2. The EEMD analysis of radio flux data at a certain frequency yields a limited number of IMFs, and these IMFs represent the intrinsic periodicities of the original data on the timescales of a few days to a solar cycle, respectively. Since this study solely focuses on investigating the rotation of radio flux, for the results of EEMD analysis, only one IMF is presented for each radio frequency, which represents the intrinsic periodicity on the timescale of a solar rotation cycle. The IMFs, representing the  signals of the solar rotation cycle of the data of 7 radio frequency fluxes, are displayed in Figure 2, while those IMFs that represent the intrinsic periodicity of the radio flux on other timescales are not presented. In order to  examine whether the EEMD method has truly extracted the signals of solar rotation cycle  from these radio flux data, particularly from the radio flux at 245 and 8800 MHz that have higher randomization, the IMFs of three typical examples of the daily radio flux at 245, 2695 and 8800 MHz  during the year 2000 are plotted again in the bottom panel of Figure 3. Meanwhile, for comparison with the original data, the radio flux  at 245, 2695 and 8800 MHz during the same time interval are displayed in the top panel of this Figure. As top panel of Figure 3 shows, both the radio flux at 2695 MHz and 8800 MHz clearly display the signal of the rotation cycle, but the randomization is a little higher for radio flux at 8800 MHz. For the radio flux at 245 MHz, visually, only a very weak modulation of the solar rotation cycle signal can be observed. This is due to the influence of the irregularities of the plasma in the corona and the Earth's upper atmosphere (Vats et al. 1997; Bhatt et al. 2017, 2018). In contrast to that, the bottom panel of this figure displays that the three IMFs all clearly exhibit the modulation due to the solar rotation cycle.
\begin{figure*}[!htbp]
\begin{center}
\includegraphics[width=1.0\textwidth]{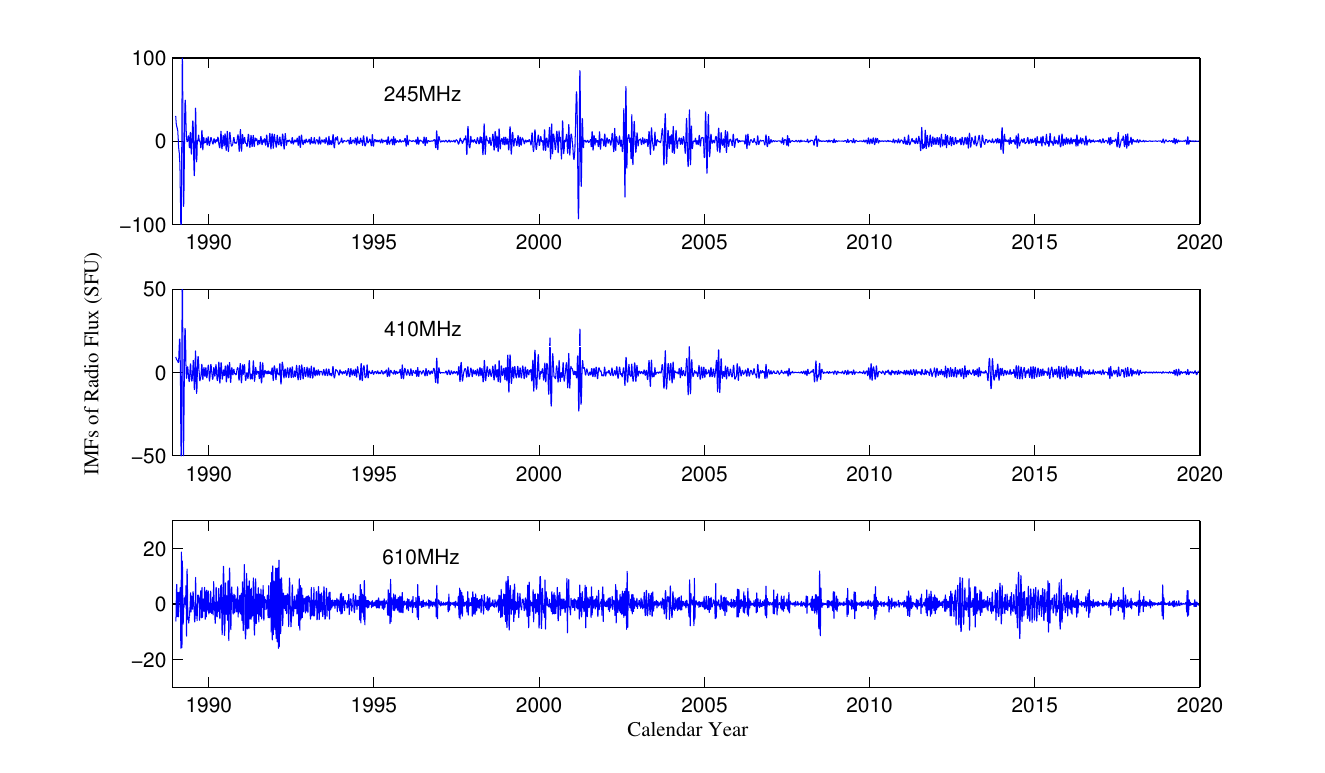}
\includegraphics[width=1.0\textwidth]{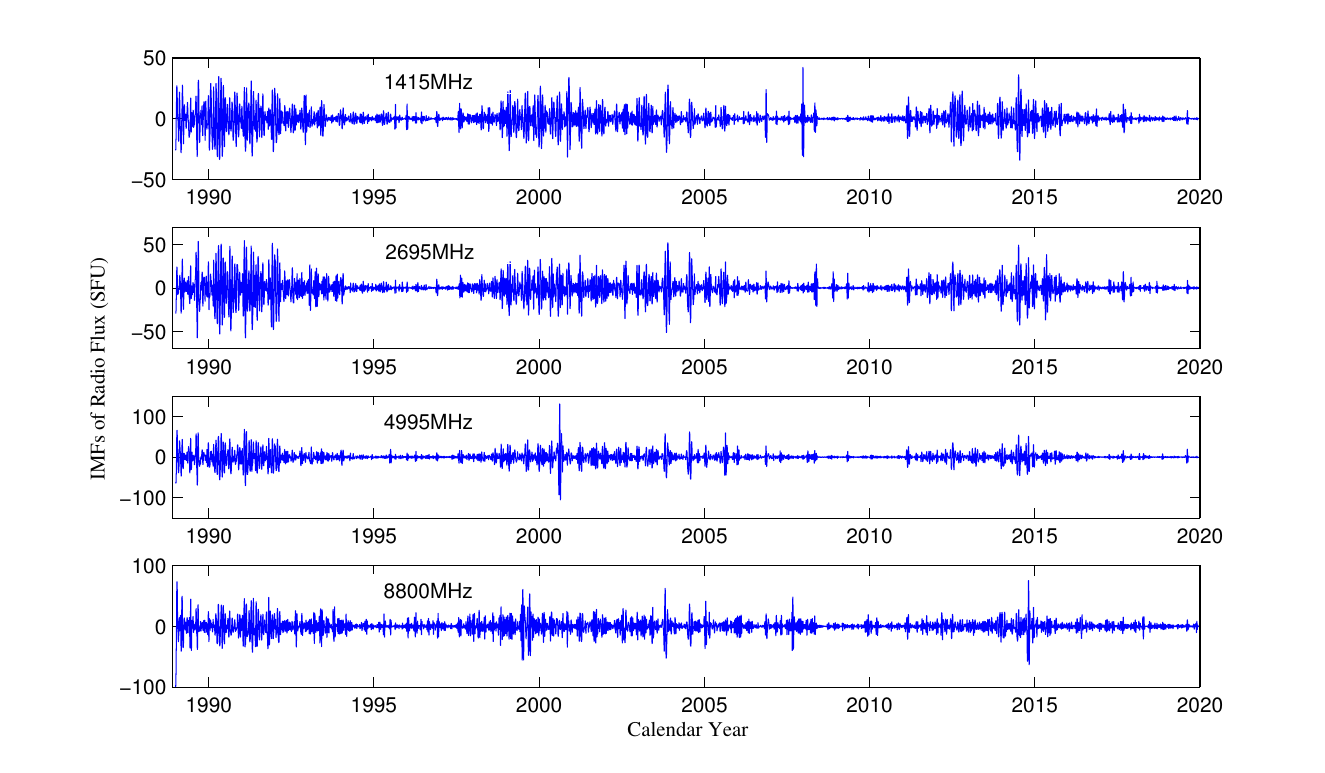}
\end{center}
\caption{Top panel: the intrinsic mode functions (IMFs) representing the signals of solar rotation cycle for solar radio flux at 245, 410, and 610 MHz.  Bottom panel: same as the top panel, but for solar radio flux at 1415, 2695, 4995, and 8800 MHz.}
\end{figure*}

\begin{figure*}[!htbp]
\begin{center}
\includegraphics[width=1.0\textwidth]{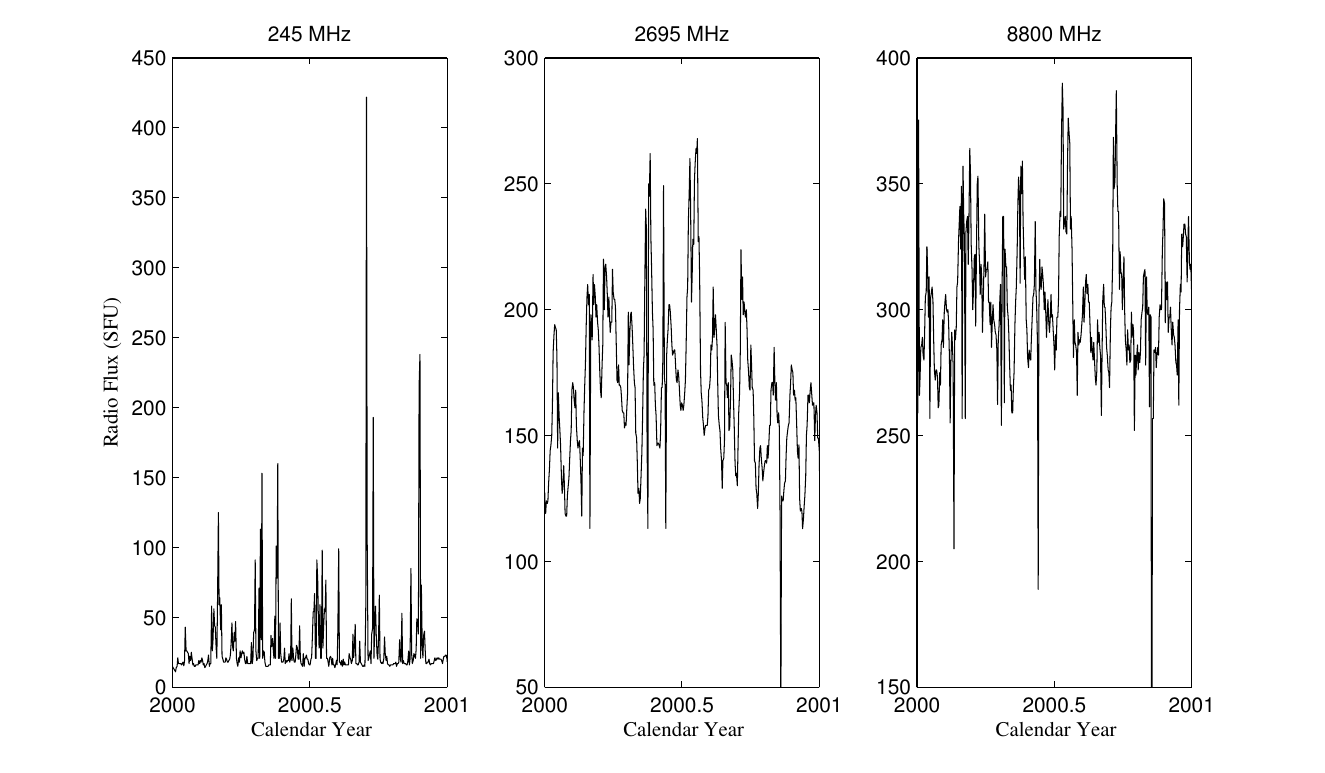}
\includegraphics[width=1.0\textwidth]{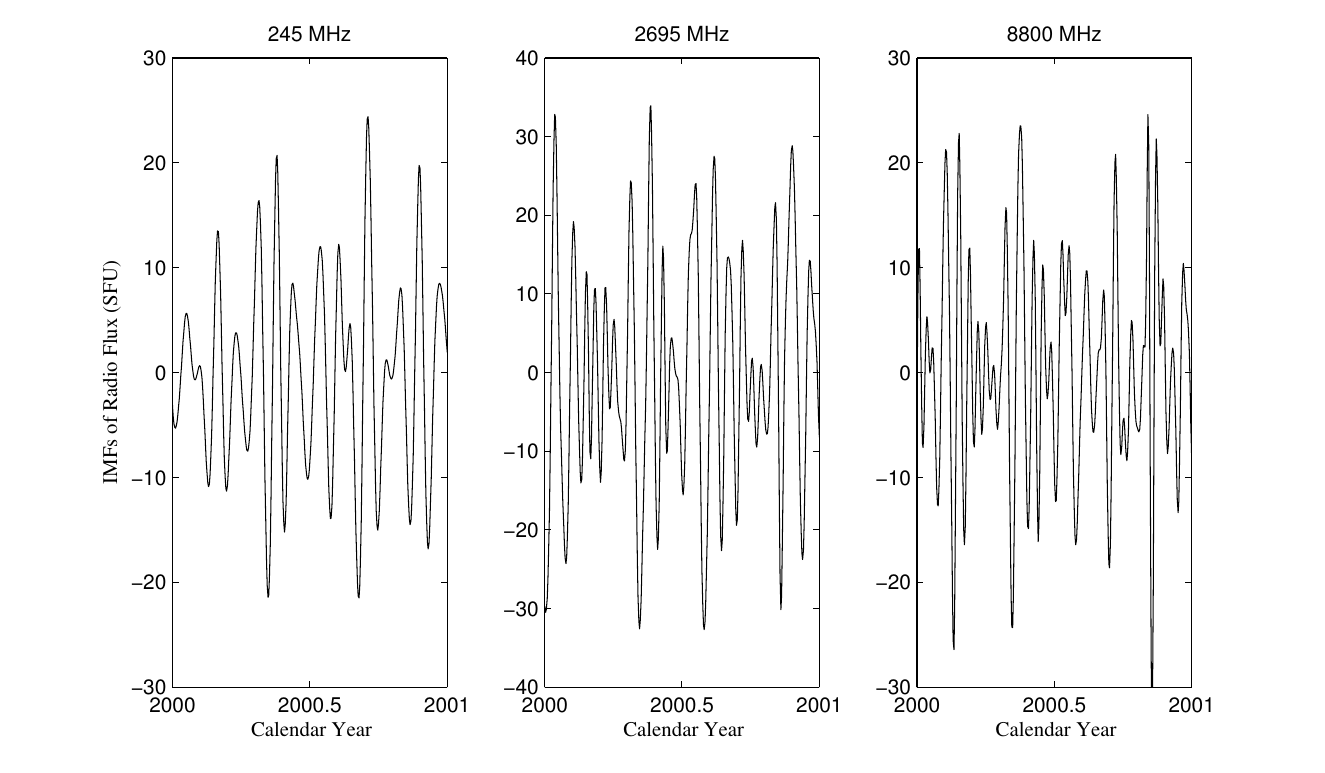}
\end{center}
\caption{Top panel: the radio flux at 245,2695 and 8800 MHz during the year 2000. Bottom panel: the IMFs representing the signals of solar rotation cycle for solar radio flux at 245, 2695 and 8800 MHz during the same time interval.}
\end{figure*}

\begin{figure*}[!htbp]
\begin{center}
\includegraphics[width=0.7\textwidth]{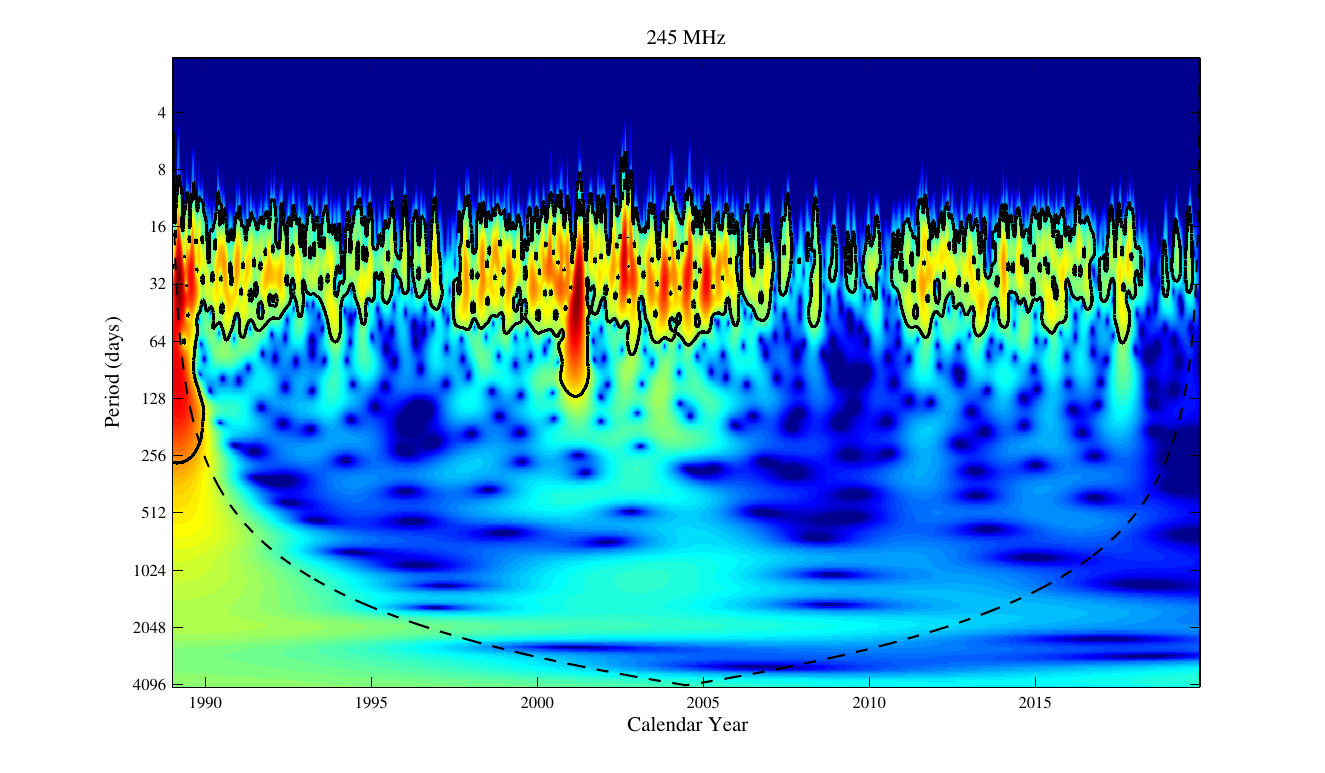}
\includegraphics[width=0.7\textwidth]{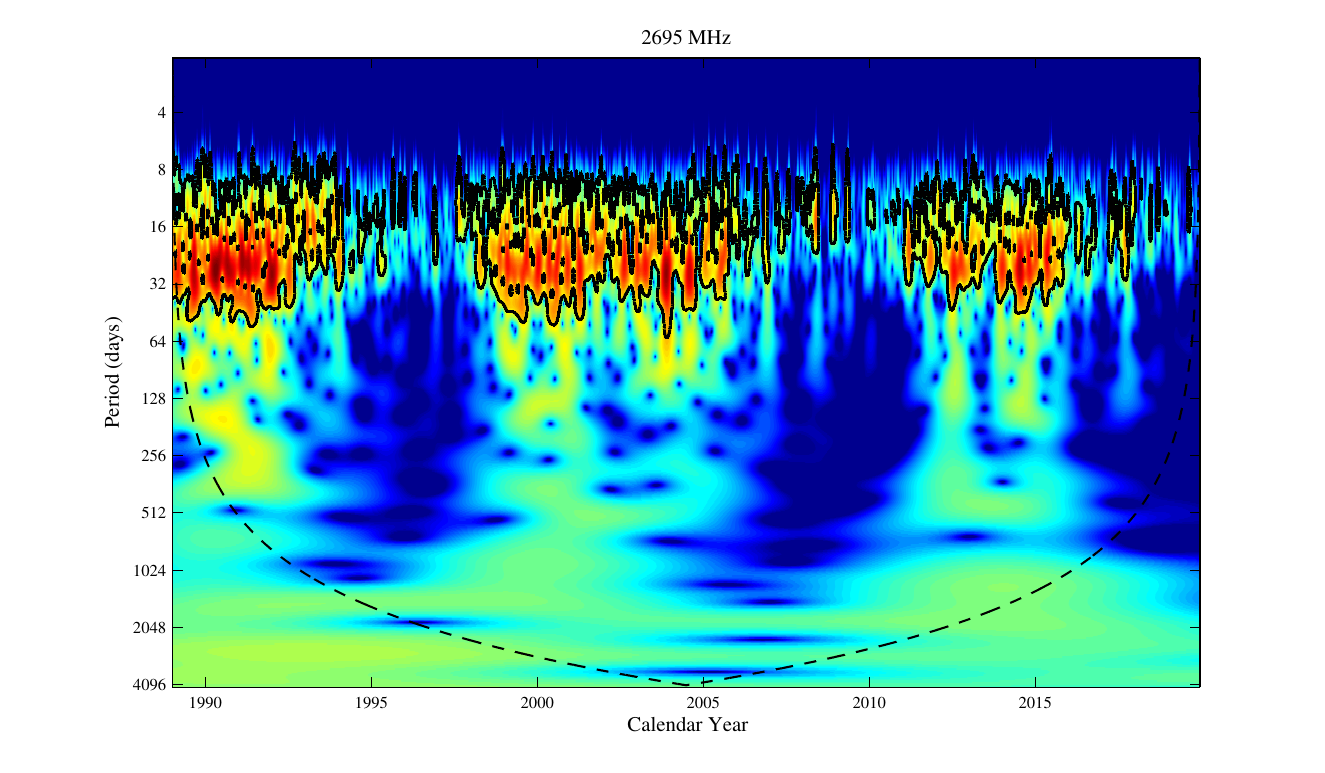}
\includegraphics[width=0.7\textwidth]{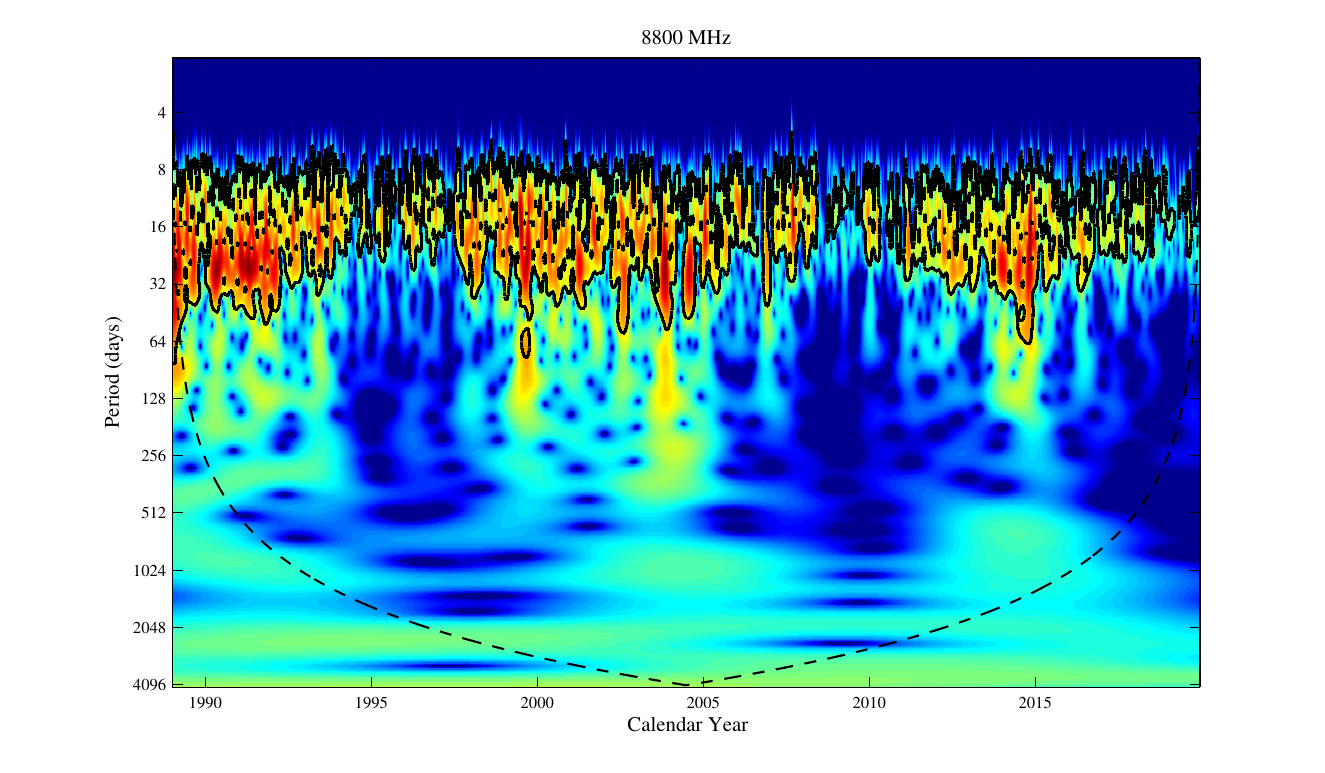}
\end{center}
\caption{Continuous wavelet power spectra of the IMFs of three typical examples of the daily radio flux at 245  (top panel), 2695 (middle panel) and 8800 (bottom panel) MHz during time interval considered.  The thick black contours represent the 99.9\% confidence level, and the black dashed line depicts the cone of influence (COI), where edge effects may distort the picture. }
\end{figure*}

\begin{figure*}[!h]
\begin{center}
\includegraphics[width=1.0\textwidth]{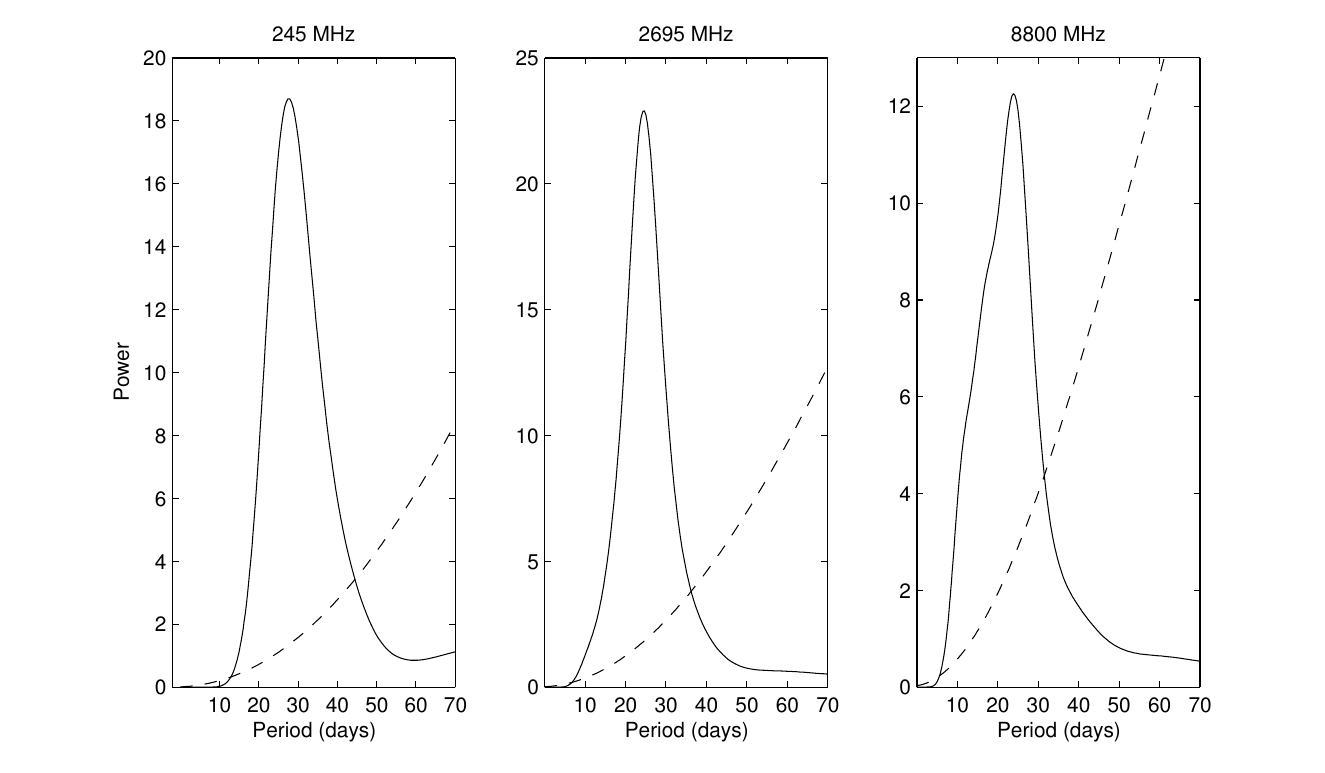}
\end{center}
\caption{Global wavelet power spectra (solid line) and the 99.9\% confidence level (dashed line) of the IMFs of three typical examples of the daily radio flux at 245 (left panel), 2695 (middle panel), and 8800 (right panel) MHz during the considered time interval. }
\end{figure*}

\begin{figure*}[!htbp]
\begin{center}
\includegraphics[width=1.0\textwidth]{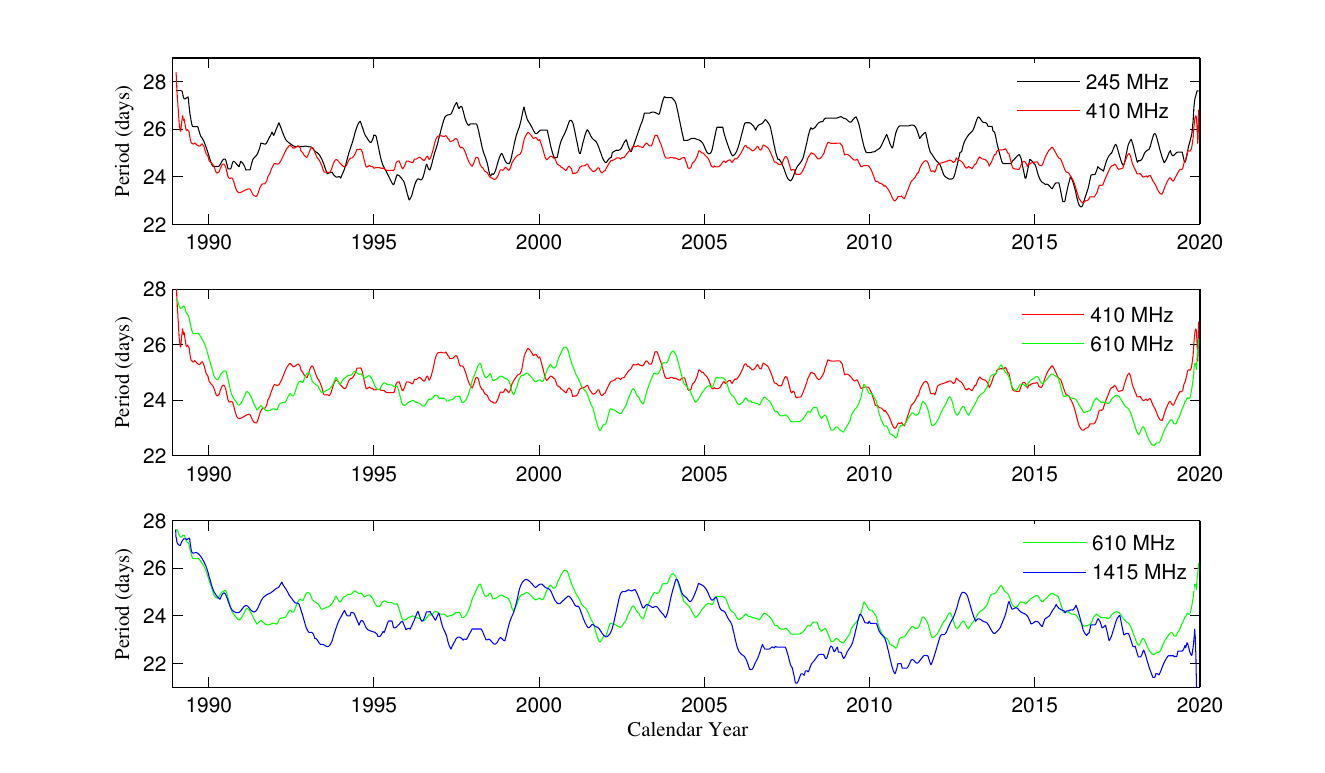}
\includegraphics[width=1.0\textwidth]{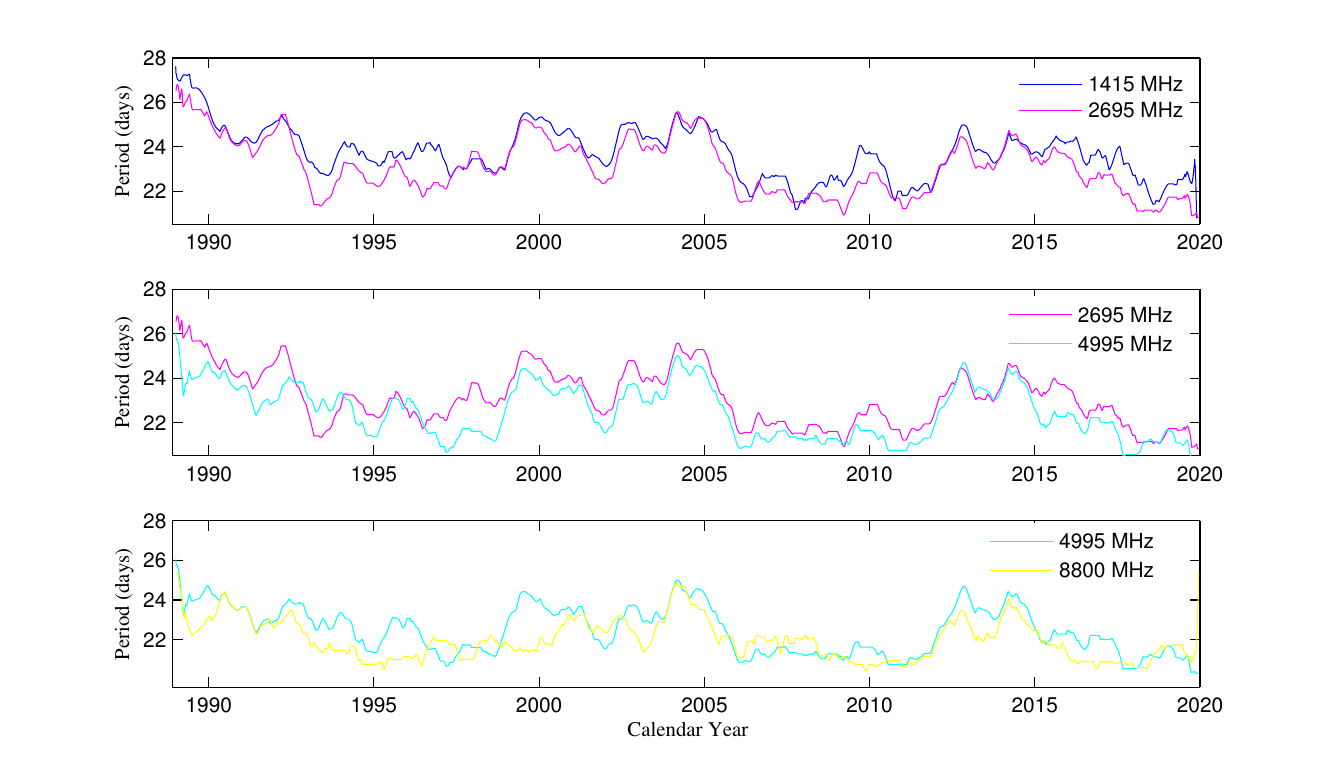}
\end{center}
\caption{The temporal variation of the rotation cycle lengths for the daily radio flux at 245, 410, 610, 1415, 2695, 4995, and 8800 MHz during the time interval considered. In order to clearly compare the rotation cycle lengths of radio flux across different frequencies, the rotation cycle lengths for only two adjacent frequencies  are displayed in a single panel. The rotation cycle lengths of different frequencies from low to high frequencies are arranged top to bottom, respectively.}
\end{figure*}

\subsection{Temporal Evolution of Radial Differential Rotation of Solar Corona}
In this study, Codes of the CWT are provided by Grinsted et al. (2004). We employ the CWT to analyze the 7 IMFs shown in Figure 2, and the analysis results only for three IMFs of typical examples of the daily radio flux at 245, 2695 and 8800 MHz are plotted in Figure 4. Here,  each panel of this figure shows the continuous wavelet power spectra of a corresponding sample IMF, highlighted by thick black contours which signify the 99.9\% confidence level. Based on the local ``components" of continuous wavelet power spectra, the global power spectra, representing the time-averaged wavelet spectrum during the entire  time interval considered, are calculated.  As a result, the global power spectra can accurately determine the dominant rotation periods for these IMFs. The global power spectra of the IMFs of three typical examples  are displayed in  Figure 5. As figure 4 shows, the continuous wavelet power spectra reveal the localized oscillatory characteristics of the sample IMFs on the timescale of one solar rotation cycle, unequivocally demonstrating the temporal variability of the rotation period for these solar radio flux data during the entire time interval considered. The global power spectra determine that the dominant synodic rotation periods for the 7 IMFs, namely those for the daily radio flux at 245, 410, 610, 1415, 2695, 4995, and 8800 MHz, are 27.12, 26.24, 25.86, 25.27, 24.62, 24.32 and 23.96 days, respectively.

The synodic rotation period, which is the apparent rotation period of the Sun observed from the Earth that orbits around the Sun, is slightly longer than the actual rotation period. This is because while the Earth revolves around the Sun, the Sun itself also rotates, and this dual rotation effect results in the observed rotation period being slightly longer.
Consequently, it is necessary  to apply a correction to the synodic rotation period and transform it into the sidereal rotation (actual) period. Following Chandra and Vats (2011), and Bhatt et al. (2017), the conversion relationship between the sidereal rotation period and the synodic rotation period is as follows:
\begin{equation}
T_{sidereal}=\frac{365.26 \times T_{synodic}}{365.26 + T_{synodic}}
\end{equation}
where the value of 365.26 represents the days in an Earth sidereal year, the $T_{sidereal}$ is the sidereal rotation period and the $T_{synodic}$ is the synodic rotation period. Through the conversion of this equation, we can obtain that the sidereal rotation periods for the daily radio flux at 245, 410, 610, 1415, 2695, 4995, and 8800 MHz are 25.25, 24.48, 24.15, 23.63, 23.07, 22.80 and 22.49 days, respectively. The synodic rotation periods of the radio flux data and their corresponding sidereal rotation periods are also presented in Table 1. As the table shows, the results clearly indicate that the rotation rate  increases as the radio emission frequency increases.
\begin{table}[!htbp]
\begin{center}
%\tablecaption{Percentage of Fake Stars Lost}
\caption[]{\centering{Synodic \& Sidereal Rotation Periods (in days) of Radio Flux (in MHz)}}%\label{Tab:publ-works}

%%Please Capitalize the First Letter of Each Notional Word in table's caption

 \begin{tabular}{clclclclcl}
  \hline\noalign{\smallskip}
   Frequencies                      & 245      & 410      & 610        & 1415      & 2695     &  4995     & 8800    \\
 synodic rotation periods           & 27.12    & 26.24    & 25.86      & 25.27     & 24.62    &  24.32    & 23.96   \\
 sidereal rotation periods          & 25.25    & 24.48    & 24.15      & 23.63     & 23.07    &  22.80    & 22.49    \\
 \hline\noalign{\smallskip}
\end{tabular}
\end{center}
\end{table}

The continuous wavelet power spectra of the typical examples depicted in Figure 4, can indicate their localized oscillatory features. Taking the top panel of this figure as an example, the local wavelet power spectra of IMF of radio flux at 245 MHz have the highest spectral power peak at a certain day on the timescale of one solar rotation cycle.  Thus, we can use these highest spectral power peaks to determine the rotation period of daily radio flux at 245 MHz at each day during the entire time interval considered. There, the results derived from the local wavelet power spectra represent the synodic rotation period. Using the Equation 2 for the calculation, we can obtain the sidereal rotation periods of the radio flux at 245 MHz at each day during the entire time interval considered. The obtained rotation periods are smoothed by the sliding  average of 1 yr, and the obtained results can indicate the variations of the rotation cycle lengths in daily radio flux at 245 MHz with solar activity. This method was also used in Xie et al. (2017) and Xiang et al. (2023) to investigate the temporal variation of the rotation of the solar corona.

In the same way, the rotation cycle lengths in daily radio flux at other frequencies considered in this study are investigated, and the obtained results are shown in Figure 6. In order to clearly compare the rotation cycle lengths of radio flux among different frequencies, each panel in Figure 6 only displays the rotation cycle lengths of radio flux for two adjacent frequencies that are considered in this study. The rotation cycle lengths of different frequencies from low to high frequencies are arranged top to bottom, respectively. As Figure 6 shows, in each panel, the rotation cycle lengths for the lower frequency of radio flux are always longer than those for the higher frequency of radio flux during almost all of the time interval considered. To sum up, over the 30-year period from 1989 to 2019, the temporal variation of the rotation cycle lengths of radio flux within the frequency range of 245-8800 MHz consistently shows a decrease in rotation cycle lengths as the frequency increases. Additionally, these values near the start and end of the time interval in each panel may have suffered from edge artifacts, therefore, these values should not indicate the true rotation cycle lengths for the radio flux data. Consequently, the parts near the start and end of the time interval considered in each panel may show different results, which should not be taken into consideration.

\section {Discussions and Conclusion}
The daily measurements of the disc-integrated solar radio flux at 245, 410, 610, 1415, 2695, 4995, and 8800 MHz during the time interval of 1989 January 1 to 2019 December 17 are used to study the temporal variation of radial differential rotation of solar corona  using the methods of EEMD and continuous wavelet transform. In this study, all the obtained rotation periods are unrelated to latitudinal differential rotation and can be regarded as averages over latitudes. Consistent with early studies, the rotation is investigated from a global perspective (Heristchi \& Mouradian 2009; Li et al. 2019; Singh et al. 2021; Xiang et al. 2023).

During the entire time interval considered, the wavelet analysis indicates that the dominant sidereal rotation periods for the daily radio flux at 245, 410, 610, 1415, 2695, 4995, and 8800 MHz are 25.25, 24.48, 24.15, 23.63, 23.07, 22.80 and 22.49 days, respectively. Based on the different mathematical methods, Vats et al. (2001),  Bhatt et al. (2017) and Singh et al. (2021) obtained the sidereal rotation periods for the daily radio flux at different frequencies within the range of 405\textendash2800 MHz during the time interval of 1997 July  to  1999 July. Although the three investigations are during the same time interval, the values of rotation periods for the same radio flux data are completely different. The sidereal rotation periods of three investigations were summarized in Figure 6 of Singh et al. (2021). By comparing the results presented in this figure with the sidereal rotation periods derived from our analysis in detail, within the frequency range of 405\textendash2800 MHz, the obtained sidereal periods and the range of their variation in this study are in agreement with Singh et al. (2021). It seems that the obtained sidereal periods in this study and Singh et al. (2021) are more accurate. On the other hand, based on analysis of the same data and the same time interval, previous studies also yielded contradictory findings regarding the variation of coronal rotation with radio flux frequency. For instance,  Vats et al. (2001) first reported that the coronal rotation period increases with increasing frequency, while Bhatt et al. (2017) and Singh et al. (2021) indicated that the coronal rotation period decreases with increasing frequency. Our findings also reveal that the dominant sidereal rotation periods for the solar radio flux within the frequency range of 245\textendash8800 MHz over the 30-year period from 1989 to 2019 show a decrease as the frequency increases. This verifies the results of Bhatt et al. (2017), and Singh et al. (2021), while our results are derived from the relationship between the average rotation period and frequency of radio flux data over a wider frequency range during a longer time interval, which should be convincing and reliable.

Wavelet analysis can not only detect the average rotation period of the corona within a certain time interval, but also investigate the temporal variation of the corona rotation period over time (Xie et al. 2017; Xiang et al. 2023). Using the method of wavelet analysis, we further study the temporal evolution of radial differential rotation of solar corona. As Figure 6 shows, the rotation cycle lengths of radio flux within the frequency range of 245-8800 MHz, over the 30-year period from 1989 to 2019, exhibit complicated temporal variations with  the progression of solar cycle. However, the rotation cycle lengths consistently  decrease with increasing frequency during  almost all of the time interval we considered.

The RSTN noon flux depicts mostly the so called "slowly varying component" on top of the quiet sun background and with occasional bursts that may occur around noontime (Kundu, 1965; Shibasaki et al. 2011; Alissandrakis 2020). Both short-lived and long-lasting transient emissions can manifest in seemingly quiet regions, hence the RSTN noon flux  can be intense and exhibit substantial temporal variations. At centrimetric wavelengths, the primary contribution to the RSTN noon flux is gyroresonant emission associated with sunspots, and a weaker contribution originates from active plage. In the absence of active regions, the emission may be associated with X-ray bright points (Keller \& Krucker 2004; Shibasaki et al. 2011; Alissandrakis et al. 2019; Alissandrakis 2020).  At relatively short decimetric wavelengths, the relative contribution of emission related to plage increases, and there is also free-free emission from coronal loops (Lantos et al. 1992; Alissandrakis 2020). In the range of even longer decimetric and metric wavelengths, the dominant long-term emitting feature is noise storm continua, which are usually believed to be derived from accelerated electrons trapped in coronal arches above active regions, with additional contributions from the lower brightness of coronal holes (Lantos et al. 1987; Lantos et al. 1992; James \& Subramanian 2018, Alissandrakis 2020). From the dominant features and generation mechanisms, the radio flux data are all directly or indirectly closely related to solar active regions. Therefore, the rotation of radio emissions is significantly influenced by their dominant features and generation mechanisms, with the most pronounced effect observed during the solar maximum. This can explain the results shown in Figures 4 and 6.

The study on the rotation of the solar upper atmosphere confirms that the rotation from the transition region to the corona is faster than that of the underlying photosphere (Vats et al. 2001; Singh et al. 2021; Sharma et al. 2020, 2021, 2023). These results and our findings can mutually validate each other. It is well known that the frozen-in effect is valid throughout the solar atmosphere. The difference between the lower and upper layers is that in the former the energy density of the plasma is higher than that of the magnetic field, which is dragged by the plasma motion (Wiegelmann et al. 2014; Bellot Rubio \& Orozco Su{\'a}rez 2019), whereas the opposite is true in the upper layer ( Alissandrakis 2020; Rao et al. 2024). Thus, the ubiquitous small-scale magnetic fields, which rotate faster than the large-scale magnetic structures (sunspots) (Xiang et al. 2014; Xu \& Gao 2016), are capable of sufficiently driving the strongly frozen-in plasma in the chromosphere and corona (Rao et al. 2024). As a result, the entire plasma atmosphere of the chromosphere and corona rotates faster than the sunspots on the solar surface (Xiang et al. 2023; Rao et al. 2024). The source regions of the radio emissions are inevitably influenced by the faster rotating plasma in the upper atmosphere, resulting in their rotation being faster than that of sunspots on the solar surface.

It is well known that radio emissions at various frequencies are derived from different altitudes, but it is difficult to determine the exact altitude of radio emissions.  What can be determined is merely that the lower frequencies originate from higher solar altitudes, while the higher frequencies from lower solar altitudes (Vats et al. 2001; Kane 2004; Alissandrakis 2020). However, in terms of the dominant feature of radio flux at each frequency, as the wavelength spans a wide range, from centrimetric to decimetric, and then to metric wavelengths, the radiation source region undoubtedly rises gradually (Lantos et al. 1987; Lantos et al. 1992; Alissandrakis 2020). Over a wide range of wavelengths, the variation in the rotation rate of radio emission with frequency appears to correspond to variation in the altitude of the  source region of radio emission. Consequently, we speculate that the regularity of rotation rates observed in radio emissions is primarily determined by altitude. To some extent, the variation in the rotation rate of radio emission with frequency can serve as an indicator of how the rotation of the solar upper atmosphere varies with altitude within a specific range.

For centrimetric wavelengths, the altitude range of source regions is very large, while the chromospheric-coronal transition region is too thin to provide any significant opacity (Shibasaki et al. 2011). Based on the radio emission mechanism, it has been determined that the RSTN noon flux at 8800 MHz originates from the transition region or the low corona (Shibasaki et al. 2011; Alissandrakis 2020). For metric wavelengths, Mercier et al. (2015) used observed data of  noise storms to investigate the relationship between the position and altitude of noise storms and the observing frequency. The authors stated that the altitudes of noise storms are approximately 1.20 $R_{\odot}$  and 1.35 $R_{\odot}$ , respectively, at frequencies of 432 MHz and 150 MHz. Meanwhile, in this wavelength range, the radio emissions  at frequencies below or equal to the plasma frequency are absorbed, whereas radiation at the second harmonic frequency, which is twice the plasma frequency, is capable of propagating (Vats et al. 2001, Shibasaki et al. 2011; Alissandrakis 2020). Consequently, based on an appropriate electron density model, the place where the radio emission is derived from can be  estimated. We select the electron density model proposed by  Aschwanden and Benz (1995) to estimate the altitude of radiation source region of radio flux at 245 MHz. The estimated results indicate that the radio emission at this frequency is derived from an altitude of approximately 1.3 $R_{\odot}$. It is important to emphasize that the calculation result of 1.3 $R_{\odot}$ is only a rough estimate, and the true radiation altitude is significantly more complex. However, combining our results with those of Mercier et al. (2015), the radiation source altitude of approximately 1.3 $R_{\odot}$ for the RSTN noon flux at 245 MHz can be considered reliable to some extent. Given this context, the obtained results of the variation in the rotation rate of radio emission with frequency indicate that the rotation of the solar upper atmosphere  at different altitudes from the low corona (or transition region) to approximately 1.3 $R_{\odot}$ exhibits complex temporal variations with the progression of the solar cycle. However, in this altitude range, over the past 30 years from 1989 to 2019, the coronal rotation consistently becomes gradually slower as the altitude increases.

Finally, our research confirms that rotation cycle signals also exist in radio emissions at both higher and lower frequencies, despite the fact that these emissions at higher or lower frequencies exhibit high randomization (Vats et al. 2001; Bhatt et al.2017, 2018), and traditional mathematical methods, such as autocorrelation, were unable to detect the rotation periods of these emissions in previous studies. The EEMD method can extract rotation cycle signals from these  highly randomized  radio emissions, and so it can be used to investigate the rotation periods for the radio emissions at higher or lower frequencies. Additionally, this work only investigates the radial differential rotation of the solar corona in the altitude range of the lower corona to approximately 1.3 $R_{\odot}$. More observation records are needed to study this topic, so as to obtain conclusive results regarding this issue in the range of the whole solar corona.

\begin{acknowledgements}
This work is supported by the National Natural Science Foundation of China (Grant Nos. 12373059 and 12203054), the Strategic Priority Research Program of Chinese Academy of Sciences (XDB 41000000), the "Yunnan Revitalization Talent Support Program" Innovation Team Project (202405AS350012), Yunnan Fundamental Research Projects (grant No. 202301AV070007), Sichuan Science and Technology Program 2023NSFSC1349, the Project Supported by the Specialized Research Fund for State Key Laboratories, and the Chinese Academy of Sciences.
\end{acknowledgements}

\end{document}